\documentclass[epj]{svjour}
\usepackage{latexsym}
\usepackage{graphics}
\usepackage[usenames,dvipsnames]{xcolor} 
\usepackage{amsmath,amssymb}

\begin{document}
\title{Hawking radiation at the zero temperature limit}
\author{Koray D\"{u}zta\c{s}\inst{1} %\and Second author\inst{2}% etc
}                     % Do not remove

\institute{Faculty of Engineering and Natural Sciences, \.{I}stanbul Okan University, 34959 Tuzla \.{I}stanbul, Turkey} %\and the second here}

\date{Received: date / Revised version: date}
\abstract{ We show that the thermal radiation derived by Hawking  can be  smoothly extended to the $T=0$ limit for Kerr black holes. The emission of the modes with $\omega > m\Omega $ comes to a halt as the surface gravity vanishes. However,  Kerr black holes smoothly continue to radiate both in bosonic and fermionic modes with $\omega < m\Omega$, at the $T=0$ limit. We derive explicit expressions for the absorption probabilities which imply that the highest rate of emission pertains to the modes with $\omega=(m\Omega)/2$, both for bosonic and fermionic cases. At the zero limit of thermal radiation, the number of emitted particles vanishes as $\omega \to  0$, which strictly differentiates it from the non-thermal radiation of soft particles by extremal Kerr black holes. We also note that the thermal radiation at the zero limit, drives the black hole away from extremality in accord with the third law and the cosmic censorship conjecture. 
%Insert your abstract here.
%
 % end of PACS codes
} %end of abstract
\maketitle

In early seventies it was proposed that certain waves should be amplified when reflected from a rotating black hole \cite{zeldo1,zeldo2,staro1,staro2}. Following this discovery, Hawking formulated the most general form of black hole radiation, which has constituted one of the milestones in black hole physics \cite{hawkingorig}. Hawking radiation has transformed our perception of black holes from being entirely ``black'' to regular objects that thermally interact with the universe. Observational verification of black hole radiation can also be expected. Though the radiation from solar mass black holes would be too faint to detect and suppressed by cosmic background radiation, it might be possible to detect gamma-ray flashes from evaporating primordial black holes, with our current technology.

In the original derivation \cite{hawkingorig}, Hawking has employed a calculation in the context of quantum field theory in curved space-times, during the formation of a black hole. In the quantum picture, the amplification of waves that are scattered by the black hole corresponds to a stimulated emission of particles. Previously, Parker and Fulling had formulated how time-dependent gravitational fields can give rise to creation of particles, using Bogoliubov transformations \cite{parker,fulling}. However, Bogoliubov transformation formalism for particle creation does not apply to stationary space-times such as Kerr family of solutions. Therefore, Hawking considered quantum fields in the time-dependent phase of a gravitationally collapsing body. He derived that there exists a steady flux of particles that reaches  the future null infinity of the Kerr space-time. The average number of particles reaching the future null infinity per unit time, depends on the energy $\omega$ and the angular momentum numbers $l,m$ as follows:
\begin{equation}
N_{\omega lm}=\frac{\Gamma_{lm}(\omega)}{\exp [2\pi (\omega - m\Omega)/\kappa] \mp 1}
\label{hawk1}
\end{equation}
where $\Omega$ is the angular velocity of the event horizon and $\kappa$ is the surface gravity. The minus and plus signs in the denominator of (\ref{hawk1}) apply to bosons and fermions respectively. $\Gamma_{lm}(\omega)$ represents the fraction of a wave with the same quantum numbers, that would be absorbed by the black hole if it were sent in from infinity. This fraction becomes negative for bosonic fields with frequency $\omega <m\Omega$, which are reflected back with a larger amplitude in the well-known process of super-radiance. However it is always positive for fermionic fields. 

In formulating the laws of black hole dynamics Bardeen, Carter and Hawking established analogies between the surface gravity of the black hole and temperature, and between the area of the black hole and entropy \cite{thermo}. Hawking radiation in the form (\ref{hawk1}) describes the thermal interaction of a black hole with the universe, with temperature $T=\kappa/2\pi$. Here, we evaluate the behaviour of Hawking radiation at the limit, the temperature --i.e. the surface gravity-- approaches zero. A similar analysis was previously carried out for extremal Reissner-Nordstr\"{o}m black holes \cite{vanzo}. An analogous effect was observed in condensed matter physics \cite{volovik}, regarding superfluids at $T=0$.

To avoid any confusion we have to distinguish between extremal black holes and the extremal limit. The number of emitted particles $N_{\omega lm}$ as a function of $\omega$ given in (\ref{hawk1}),  is undefined at the point $\kappa=0$. Therefore extremal black holes cannot be associated with thermal radiation. Recently it was  shown that there exists a non-thermal radiation with no defined temperature for extremal Reissner-Nordstr\"{o}m and Kerr black holes \cite{goodrn,goodkerr}. In both cases, the number of particles emitted by extremal black holes diverges as the energy of the particles approaches zero; i.e. extremal black holes emit soft particles in a non-thermal fashion. However the limit $\kappa \to 0$ of (\ref{hawk1}) is well defined, which will be evaluated in this work.
  
The essential difference between Hawking radiation and the spontaneous emissions previously derived by Zel'dovich and Starobinsky \cite{zeldo1,zeldo2,staro1,staro2} is that it allows the emission of the modes with $\omega > m\Omega$. The probability of observing these modes smoothly decreases as the surface gravity $\kappa$ decreases. At the limit $\kappa \to 0$, the exponential term in the denominator of (\ref{hawk1}) diverges  both for fermionic and bosonic modes with $\omega >m\Omega$ .  For $\omega > m\Omega$, the thermal radiation comes to a halt at the zero limit in accord with classical electromagnetic radiation.

However, the intuition that there cannot be a thermal radiation at zero temperature limit, leads us astray. Consider the modes with $\omega <m\Omega$. In this case, the exponential term in the denominator becomes zero  at the zero temperature limit. For bosonic modes with $\omega <m\Omega$, the zero temperature limit is given by:
\begin{equation}
\lim_{T \to 0} N_{\omega lm}^{\rm{B}}=\frac{\Gamma_{l,m}(\omega)}{- 1}=\vert \Gamma_{lm}(\omega) \vert
\label{zeroboson}
\end{equation}
where the superscript $\rm{B}$ refers to bosons. Note that $\Gamma_{l,m}(\omega)$ is negative for bosonic modes with $\omega <m\Omega$. For fermionic modes with $\omega <m\Omega$, the zero temperature limit is:
\begin{equation}
\lim_{T \to 0} N_{\omega lm}^{\rm{F}}=\frac{\Gamma_{lm}(\omega)}{ 1}= \Gamma_{lm}(\omega) 
\label{zerofermion}
\end{equation}
where the superscript $\rm{F}$ refers to fermions. Note that $\Gamma_{lm}(\omega)$ remains positive for fermionic modes with $\omega <m\Omega$. 

The fraction of a wave that is absorbed by the black hole is conventionally called the absorption probability. Though the phrase ``probability'' is inappropriate for a quantity that can acquire a negative value (see \cite{qu1} for an elucidative discussion), we shall adapt the conventional term ``absorption probability'' here. The absorption probabilities for bosonic and fermionic fields scattering from Kerr black holes were derived by Page \cite{page}. The absorption probabilities determine the flux of radiated particles at the asymptotically flat infinity, both for non-extremal ($\kappa>o$) and extremal ($\kappa=o$) black holes. For bosonic fields Page derived that:
\begin{eqnarray}
\Gamma^{\rm{B}}&=&\left[ \frac{(l-s)!(l+s)!}{(2l)!(2l+1)!!}\right]^2 \prod_{n=1}^{l}\left[1+\left(\frac{\omega - m\Omega}{n\kappa}\right)^2 \right] \nonumber \\
&\times& 2 \left(\frac{\omega - m\Omega}{\kappa}\right)  \left(\frac{A\kappa}{2\pi}\omega \right)^{2l+1}
\label{probpage}
\end{eqnarray}
where $s=0,1,2$ denotes the spin of the bosonic field, and $A$ is the surface area of the black hole. The absorption probability becomes negative for $\omega < m\Omega$, which justifies that super-radiance occurs for these modes. The highest absorption probability pertains to the modes with $l=s$. Page simplified the expression (\ref{probpage}) for small $\omega$, but we do not pre-assume that $\omega$ is small in this derivation. For $l=s=1$, (\ref{probpage}) takes the form:
\begin{equation}
\Gamma^{\rm{B}}_{(l=s=1)}=\frac{2}{9}\left( \frac{A\omega}{2\pi} \right)^3(\omega -m\Omega) \left[ \kappa^2 + (\omega -m\Omega)^2 \right]
\end{equation}
Note that the spin $s$ only appears in the calculation of the coefficient in the square brackets in (\ref{probpage}).  therefore for spin-0 fields the absorption probability satisfies:
\begin{equation}
\Gamma^{\rm{B}}_{(l=1,s=0)}=\frac{1}{4} \Gamma^{\rm{B}}_{(l=s=1)}
\label{l1s0}
\end{equation}
The modes with $l=0,m=0$ are irrelevant for our calculation, since we impose $\omega - m\Omega <0$ to have a positive number of particles reaching the null infinity. The absorption probability in the extremal limit for the modes with $l=1,m=1$  is given by:
\begin{equation}
\Gamma^{\rm{B}}_{(l=s=1)}=\frac{2}{9}\left( \frac{A}{2\pi} \right)^3 [\omega(\omega -m\Omega)]^3
\label{extls1}
\end{equation}
All the modes with $l=1,m=1$ have absorption probabilities equal to a fraction of (\ref{extls1}) as we have derived for 
$(l=1,s=0)$ in (\ref{l1s0}). The average number of particles emitted by Kerr black holes at the extremal limit  $(\kappa \to 0 ; T \to 0)$ via  Hawking radiation in the modes $(l=s=1)$ is given by:
\begin{equation}
N_{l=m=1}^{\rm{B}}(\omega)=C_{\rm{lsm}}\left( \frac{A}{2\pi} \right)^3 \vert [\omega(\omega -m\Omega)]^3 \vert
\label{hawkl1s1}
\end{equation}
Note that, $C_{\rm{lsm}}=2/36$ for $l=m=1,s=0$, $C_{\rm{lsm}}=2/9$ for $l=m=1,s=1$. 

The number of particles given as a function of $\omega$ in (\ref{hawkl1s1}), acquires its maximum value at:
\begin{equation}
\omega_{\rm{max}}=m\Omega/2
\end{equation}
For the case of spin-2, the absorption probability (\ref{probpage}) takes the form:
\begin{eqnarray}
\Gamma^{\rm{B}}_{(l=s=2)}&=&\left( \frac{2}{225}\right)  (\omega - m\Omega)  \left(\frac{A\omega}{2\pi} \right)^{5}   \nonumber \\
&\times& \left( \kappa^4+ \frac{5\kappa^2}{4} (\omega - m\Omega)^2 + \frac{(\omega - m\Omega)^4}{4}  \right)  
\label{probgen}
\end{eqnarray}
which we have also derived in a recent work \cite{spin2}. For extremal black holes (\ref{probgen}) reduces to:
\begin{equation}
\Gamma^{\rm{B}}_{(l=s=2)}(\omega) =\left( \frac{1}{450}\right) \left(\frac{A}{2\pi} \right)^{5} [\omega (\omega - m\Omega)]^5
\label{probext}
\end{equation}
which implies that the average number of particles emitted by extremal black holes via Hawking radiation with $m=2$ is given by:
\begin{equation}
N_{l=m=2}^{\rm{B}}(\omega)=C_{\rm{lsm}}\left( \frac{A}{2\pi} \right)^5 \vert [\omega(\omega -m\Omega)]^5 \vert
\label{hawkl2m2}
\end{equation}
where $C_{\rm{lsm}}$ becomes maximum at $l=s=m=2$. Again the number of particles expressed as a function of $\omega$, attains its maximum value at $\omega=m\Omega/2$.

For fermionic fields the general form of the absorption probability is given by \cite{page}:
\begin{eqnarray}
\Gamma^{\rm{F}}&=&\left[ \frac{(l-s)!(l+s)!}{(2l)!(2l+1)!!}\right]^2 \left(\frac{A\kappa}{2\pi}\omega \right)^{2l+1} \nonumber \\ 
&\times & \prod_{n=1}^{(l+1/2)}\left[1+\left(\frac{\omega - m\Omega}{n\kappa -\frac{1}{2}\kappa}\right)^2 \right]
\label{pageprobfer}
\end{eqnarray}
For $l=s=1/2$ the absorption probability takes the form:
\begin{equation}
\Gamma^{\rm{F}}_{(l=s=1/2)}=\frac{1}{4} \left( \frac{A \omega}{2\pi} \right)^2 [\kappa^2 +4(\omega - m\Omega)^2]
\label{probls1half}
\end{equation}
Note that the absorption probability  remains positive for $\omega < m\Omega$. At the extremal limit limit the absorption probability (\ref{probls1half}) reduces to:
\begin{equation}
\Gamma^{\rm{F}}_{(l=s=1/2)}= \left( \frac{A }{2\pi} \right)^2 [\omega(\omega - m\Omega)]^2
\label{probls1halfext}
\end{equation}
The probability at the extremal limit given in (\ref{probls1halfext}) determines the average number of particles with  $m=1/2$,  emitted via Hawking radiation by extremal black holes.
\begin{equation}
N_{l=m=1/2}^{\rm{F}}(\omega)=\Gamma^{\rm{F}}_{(l=s=1/2)}
\end{equation}
As in the bosonic case, the average number of particles attains its maximum value at $\omega=m\Omega /2$. We can also evaluate the absorption probability for $s=3/2$.
\begin{equation}
\Gamma= \frac{1}{64} \left(\frac{A \omega}{2\pi} \right)^4 \left[\kappa^4 +\frac{40\kappa^2(\omega - m\Omega)^2 }{9}+ \frac{16(\omega - m\Omega)^4}{9}\right]
\label{probthreehalves}
\end{equation}
which was also derived in \cite{threehalves}. In the extremal limit, (\ref{probthreehalves}) reduces to
\begin{equation}
\Gamma= \frac{1}{36} \left(\frac{A }{2\pi} \right)^4 \left[\omega (\omega - m\Omega) \right]^4
\label{probthreehalvesext}
\end{equation}
The absorption probability for extremal black holes also determines the average number of fermionic particles   emitted via Hawking radiation as described in (\ref{zerofermion}). 

The average number of particles emitted by extremal black holes in bosonic and fermionic modes can be compactly written in the form:
\begin{eqnarray}
N_{l,m}(\omega)&=& C_{\rm{lsm}} \left(\frac{A }{2\pi} \right)^{2m+1}  \left |[\omega (\omega - m\Omega)]^{2m+1}\right| \nonumber \\
&& \qquad m= \{ (1/2),1,(3/2),2 \}
\label{ngeneric}
\end{eqnarray}
where $C_{\rm{lsm}}$ becomes maximum for $l=m=s$. Both for bosonic and fermionic modes $N_{l,m}(\omega)$ attains its maximum value at $\omega=m\Omega /2$. Note that the derivation pre-assumes $\omega < m\Omega$ to ensure that the exponential term in the denominator does not diverge at $\kappa =0$. 

Recently, it was shown that extremal Kerr and Reissner-Nordstr\"{o}m black holes exhibit a non-thermal radiation with no defined temperature \cite{goodrn,goodkerr}. The number of emitted particles diverges as $\omega$ approaches zero, hence soft particles are emitted. On the other hand, the expression (\ref{ngeneric}) implies that the thermal radiation at the zero temperature limit becomes maximum for $\omega=m\Omega/2$ and ceases as $\omega$ approaches zero. The thermal radiation  at the limit $\kappa \to 0$ and the non-thermal radiation at $\kappa=0$ behave essentially differently. The fact that these two cases cannot be reconciled, lends credence to the argument that extremal black
holes cannot be considered as the limit of nearly extremal black holes. (See e.g. \cite{carroll})

Let us also examine the consequences of emitting particles with energy $\omega < m\Omega$. The mass $M$ and angular momentum $J$ parameters of an extremal Kerr black hole satisfy $M^2 -J=0$. If the angular momentum parameter increases beyond the extremal limit, the event horizon which disables the causal contact of the singularity with distant observers, ceases to exist. In that case, the space-time parameters describe a naked singularity.  In its weak form, the cosmic censorship conjecture proposed by Penrose asserts that all curvature singularities should be hidden behind event horizons of black holes \cite{ccc}. The amount mass and angular momentum carried away by the emitted
particles are related by
\begin{equation}
\delta M =\frac{m}{\omega} \delta J
\label{deltamdeltaj}
\end{equation}

The emission of the modes with energies  $\omega < m\Omega$ slightly decreases the mass parameter of the space-time. However, it leads to a larger decrease in the angular momentum parameter, as implied by (\ref{deltamdeltaj}). The final parameters of the space-time satisfy:
\begin{equation}
(M-\delta M)^2-(J-\delta J)=(\delta M)^2 -2M\delta M -\frac{m}{\omega}\delta M
\label{extfin}
\end{equation}
Note that $\Omega=1/2M$ for extremal black holes and $\omega < m\Omega$ implies that the expression in (\ref{extfin}) is positive. The emission of particles with energies $\omega < m\Omega$, drive the black hole away from extremality to a non-extremal state, in accord with the cosmic censorship conjecture. The black hole which was initially in extremal state with $\kappa=0$, acquires a surface gravity or equivalently a temperature. For nearly extremal black holes the surface gravity $\kappa$ is small, and the average number of emitted particles  as a function of the frequency, attains its maximum value around $\omega=m\Omega /2$, which would at most be shifted to first order. Hawking radiation also drives nearly-extremal black holes away from extremality in accord with the third law of black hole dynamics \cite{israel1}, and cosmic censorship. We have previously alluded to this fact in \cite{ha}.

Hawking radiation originally derived in the form (\ref{hawk1}) behaves smoothly at the limit $\kappa=0$. The thermal radiation by  Kerr black holes can be smoothly extended to the extremal limit with $T=0$. Here, the emission is purely due to the Hawking effect which should not be confused with super-radiance which smoothly  continues both at the extremal limit ($T\to 0$), and for extremal black holes ($T=0$) \cite{dejan1,dejan2}. At this limit the number of the particles emitted with energies $\omega > m\Omega$, decreases to zero. However,  Kerr black holes at the extremal limit continue to emit particles with energies $\omega < m\Omega$, both in bosonic and fermionic modes. The average number of particles emitted by  Kerr black holes at the extremal limit is derived in (\ref{ngeneric}) as a function of $\omega$, which attains its maximum value at $\omega=m\Omega /2$. The behaviour at the zero temperature limit, does not coincide
with the non-thermal radiation of soft particles by
extremal black holes.  The emission of these modes drives the black hole away from extremality, which leads the black hole to acquire a positive temperature. Radiation at the zero temperature limit is a natural mechanism reinforcing the stability of the event horizon in accord with the third law of black hole dynamics and the cosmic censorship conjecture.

%
% BibTeX users please use
% \bibliographystyle{}
% \bibliography{}

\end{document}